\documentclass[twocolumn]{aa}
\usepackage{graphicx,color}
\usepackage{txfonts}
\usepackage{amsmath}
\usepackage{textcomp}
\usepackage{gensymb}
\usepackage{tabularx}
\usepackage{rotating}
\usepackage{listings}
\usepackage{float}
\usepackage{hyperref}

\makeatletter
\renewcommand*\aa@pageof{, page \thepage{} of \pageref*{LastPage}}
\makeatother
\begin{document} 

\title{Constraining the period of the ringed secondary companion to the young star J1407 with photographic plates}

   \author{R. T. Mentel \inst{1}$^,$\inst{2} [0000-0002-5178-8083]
          \and
          M. A. Kenworthy\inst{1} [0000-0002-7064-8270]
          \and 
          D. A. Cameron\inst{3}
          \and 
          E. L. Scott\inst{4}
          \and
          S. N. Mellon \inst{3} [0000-0003-3405-2864]
          \and
         R. Hudec\inst{5}$^,$\inst{6}
         \and
         J. L. Birkby\inst{7}$^,$\inst{8}
          \and 
         E. E. Mamajek \inst{9}$^,$\inst{3} [0000-0003-2008-1488]
          \and
         A. Schrimpf \inst{2}  [0000-0001-5860-4139]
          \and
         D. E. Reichart \inst{10}
          \and
         J. B. Haislip \inst{10}
          \and
         V. V. Kouprianov \inst{10}
          \and
          F.-J. Hambsch \inst{11}$^,$\inst{12}
          \and
         T.-G. Tan \inst{11}$^,$\inst{13} [0000-0001-5603-6895]
          \and
         K. Hills \inst{11}
          \and
         J. E. Grindlay \inst{8}
          }

   \institute{
   Leiden Observatory, Leiden University, PO Box 9513, NL-2300 RA Leiden, the Netherlands\\
   \email{mentel@strw.leidenuniv.nl}
   \and
    Department of Physics, Philipps-Universit{\"a}t Marburg, Renthof 5, 35032 Marburg, Germany
	\and
    Department of Physics \& Astronomy, University of Rochester, 500 Wilson Blvd., Rochester, NY 14627, USA
    \and
    Chinese Academy of Sciences, 52 Sanlihe Road, Xicheng District, 100864 Beijing, China
    \and 
	Faculty of Electrical Engineering, Czech Technical University, Zikova 1903/4, 166 36 Prague, Czech Republic
	\and
	Kazan Federal University, 18 Kremlyovskaya Street, 
420008 Kazan, Russian Federation
	\and
    Anton Pannekoek Institute for Astronomy, University of Amsterdam, Science Park 904, NL-1098 XH Amsterdam, the Netherlands
    \and
    Harvard-Smithsonian Center for Astrophysics, 60 Garden Street, Cambridge, MA 02138, USA
    \and
	Jet Propulsion Laboratory, California Institute of Technology, M/S 321-100, 4800 Oak Grove Drive, Pasadena, CA 91109, USA
    \and
Department of Physics and Astronomy, University of North Carolina at Chapel Hill, Campus Box 3255, Chapel Hill, NC 27599, USA
    \and
    American Association of Variable Star Observers (AAVSO)
    \and
    Center for Backyard Astrophysics (Antwerp), American Association of Variable Star Observers (AAVSO), Vereniging Voor Sterrenkunde (VVS), ROAD Observatory, Oude Bleken 12, B-2400 Mol, Belgium
    \and
    Perth Exoplanet Survey Telescope
}

   \date{Received 03 August 2018; accepted 04 September 2018}

  \abstract
   {The 16 Myr old star 1SWASP J140747.93-394542.6 (V1400 Cen) underwent a series of complex eclipses in May 2007, interpreted as the transit of a giant Hill sphere filling debris ring system around a secondary companion, J1407b.
   No other eclipses have since been detected, although other measurements have constrained but not uniquely determined the orbital period of J1407b.
   Finding another eclipse towards J1407 will help determine the orbital period of the system, the geometry of the proposed ring system and enable planning of further observations to characterize the material within these putative rings.}
   {We carry out a search for other eclipses in photometric data of J1407 with the aim of constraining the orbital period of J1407b.}
   {We present photometry from archival photographic plates from the Harvard DASCH survey, and Bamberg and Sonneberg Observatories, in order to place additional constraints on the orbital period of J1407b by searching for other dimming and eclipse events.
   Using a visual inspection of all 387 plates and a period-folding algorithm we performed a search for other eclipses in these data sets.}
   {We find no other deep eclipses in the data spanning from 1890 to 1990, nor in recent time-series photometry from 2012-2018.}
   {We rule out a large fraction of putative orbital periods for J1407b from 5 to 20 years.
   These limits are still marginally consistent with a large Hill sphere filling ring system surrounding a brown dwarf companion in a bound elliptical orbit about J1407.
   Issues with the stability of any rings combined with the lack of detection of another eclipse, suggests that J1407b may not be bound to J1407.}
%
\titlerunning{J1407 photometry}
\authorrunning{Mentel et al.}
\setlength{\headheight}{20pt}
   \maketitle
   
\section{Introduction}

All the gas giants in the solar system have multiple moons in orbit around them, 
and their nature gives insight into the earliest processes of planet and moon formation in the first $10-50$ Myr of the solar system \citep[e.g. see reviews by ][]{Armitage11,Kley12}.
Planets presumably form through accretion within circumstellar disks, whereas most moons orbiting giant planets are thought to form from accretion of material within circumplanetary disks. 
The circumplanetary disk subsequently accretes onto the planet or into moons \citep{Canup02,Magni04,Ward10}.
Exomoons are expected to be common around giant exoplanets, and some projects are already under way searching for the signs of large satellites.
The Hunt for Exomoons with Kepler (HEK) survey has searched for satellites around exoplanets in the Kepler data since 2011\citep{Kipping12}, yielding only one marginal candidate for a potential exomoon\citep{Teachey18} and placing strong upper limits on large moons orbiting short period $(P\,<\,20\,d)$ exoplanets \citep{Kipping18}.
The possibility of tidal heating of exomoons in Laplace resonances \citep{Peters13}, combined with heating from the central planet, suggest that Habitable Zones can exist around giant exoplanets \citep{Heller14b}.
The direct detection of an exo-circumplanetary disk presents the opportunity to study both the dynamics and composition of the forming planet.

The young star 1SWASP J140747.93-394542.6, recently given the variable star designation V1400~Cen\footnote{V1400 Cen was assigned by \citet{Samus17} in the General Catalogue of Variable Stars version 5.1.} (and hereafter referred to as J1407), is a probable kinematic member of the Sco-Cen OB Association \citep{Mamajek12,Pecaut16}.
J1407 is a very active, fast-rotating ($\sim$3 day) Li-rich K5 pre-main sequence star with estimated mass of $\sim 0.9\,M_\odot$.
The Gaia Data Release 2 \citep{Gaia16,Gaia18} provides the first precise trigonometric parallax for V1400 Cen of $\varpi$ = 7.1835\,$\pm$\,0.0447 mas, consistent with $d$ = 139.2\,$\pm$\,0.8 pc.
\footnote{This is in good agreement with earlier kinematic distance estimates \citep[128\,$\pm$\,13 pc;][]{Mamajek12}, and the mean distance to the UCL subgroup of Sco-Cen \citep[142\,pc;][]{deZeeuw99}.
A calculation with the BANYAN $\Sigma$ tool\citep{Gagne18} using these new parameters yielded a near-certain probability of UCL membership of 96.1$\%$.}
We have revised the distance, brightness, age, and mass of the star using these new parameters and provide them in Table \ref{tab:stellar_params}.
A complex sequence of deep eclipses of J1407 in time-series photometry
data from May 2007 was reported by \citet{Mamajek12} and interpreted being as due to a large ring-like structure of debris, some 200 times larger than Saturn's rings \citep{Mamajek12,vanWerkhoven14,Kenworthy15b}.
The photometric variability during the transit ranges on timescales from hours through to months, suggesting a large coherent structure of the order of $\sim$1 AU in diameter.
The leading hypothesis is of an unseen secondary companion providing the gravity well in which this structure is resident.

To date, a search for the secondary companion through direct imaging have placed upper limits on the orbital period and mass \citep{Kenworthy15}.
Subsequent photometric monitoring has not detected a second eclipse, and photometry from wide field surveys using large format imaging cameras from the past 15 years shows no other eclipse.

Over the past decade, archival photographic plates from DASCH (USA) as well as the Sonneberg Observatory and Bamberg Observatory in Germany, have been digitized and light curves for all detected objects are now publicly available.
J1407 is detected on about 1000 photographic plates that cover a baseline of over one century.
In this paper we present the photometric analysis of these photographic plates in search of one or more additional eclipses, and include the latest photometry from several ongoing surveys.
The photographic plate photometry of J1407 is presented for the first time.

In Section~\ref{sec:obs} we describe the photographic plate archives used for the search.
In Section~\ref{sect_analysis} we describe the search and analysis of the photographic plates.
We describe the code we use to infer limits on the orbital period of the companion from the epochs of observation.
In Section~\ref{sec:conc} we place the results in context of other measurements of the J1407 system.

\begin{table}
\centering
\begin{tabular}{ccc}
\hline
Property & Value & Source \\ 
\hline
$\alpha(J2000)$ & 14:07:47.93 & (1) \\
$\delta(J2000)$ & -39:45:42.7 & (1) \\
Spectral Type & K5 IV(e)Li & (2) \\
Distance & 139.2 $\pm$ 0.8 pc & (3) \\
$T_{eff}$ & $4500^{+100}_{-200}$K & (2) \\
$R$ & 0.96 $\pm\ 0.15R_\odot$ & (2) \\
Age & 21.38 +4.3 -7.6 Myr & (3) \\
Mass & 0.95 $\pm\ 0.1 M_\odot$ & (3)\\
RV & 14.6 $\pm$ 0.4 km/s & (4)\\
\hline
\end{tabular}
\label{tab:stellar_params}
\caption{Stellar Parameters of J1407 from \citep{Zacharias10} (1), \citep{Mamajek12} (2), newly calculated with data from the Gaia Data Release 2 (3), and from \citet{Kenworthy15} (4).}
\end{table}

\section{Observations}\label{sec:obs}

We describe the sources of photometric data used in our analysis, along with Table~\ref{tab:photographs} that details the photometric observations from photographic plates.

\begin{table*}[t]
\centering
\caption{Summary of new photometry extracted from the photographic plates detailed in this paper.}
\label{tab:photographs}
\begin{tabular}{lccc}
\hline
Name  & Total photometric points & Total used in current analysis  & Mean Error [mag] \\
\hline
DASCH & 989 & 636 & 0.15 \\
Sonneberg & 103 & 103 & 0.30 \\
Bamberg & 171 & 109 & 0.23 \\
SuperWASP & 28918 & 28918 & 0.02 \\
AAVSO & 4342 & 4243 & 0.026 \\
KELT & 5534 & 5534 & 0.038 \\
PROMPT & 2013 & 2013 & 0.018 \\
\hline
\end{tabular}

\end{table*}

\subsection{Photographic plates from DASCH}
The Harvard College Observatory undertook an all-sky photographic survey of both the southern and northern hemisphere with various telescopes between 1885 and 1993, producing $\sim 500,000$ photographic plates, which are being digitized by the ongoing project `Digital Access to a Sky Century @Harvard' \citep[DASCH;][]{Tang2013}.

From the `b'-series, 45 were recorded between 1890 and 1950 with a 8'' Bache Doublet, Voigtlander at Mount Harvard in Colorado, Arequipa in Peru and Bloemfontein in South Africa.
Another 350 plates from the `am'-series were taken between 1901 and 1926 using a 1'' and 1.5'' Cooke Lens in Arequipa, Peru, and Bloemfontein, SA.
In 1909 and 1910, four plates from the `ak'-series were recorded with a 1.5'' Cooke `Long Focus' in Hanover, SA. 
Another 138 plates in the `mf'-series were taken between 1919 and 1949 with a 10'' Metcalf triplet in Arequipa, Peru and Bloemfontein, SA.
A set of 112 plates in the `ax'-series were recorded between 1924 and 1948 using a 3'' Ross-Tessar lens  in Arequipa, Peru and Bloemfontein, SA and 215 plates in the `rb'-series were recorded between 1929 and 1951 with a 3'' Ross-Fecker in Bloemfontein, SA.
An additional 125 plates in the `dsr'-, `dsy'- and `dsb'-series  were taken with three Damon telescopes equipped with red filter (9 plates), a yellow filter (9 plates and a blue filter (106 plates) in 1970 and 1971 in Bloemfontein, SA (dsr, dsy and dsb) as well as between 1978 and 1989 on Cerro Tololo, Chile and Mount John, New Zealand (dsb).

For our analysis (see Figure \ref{fig_phot}) we used 989 photographic plates from the DASCH archive, observed between 1890 and 1989.
These are made up of 602 plates whose photometry we acquired via the DASCH pipeline, and 387 plates which were not processed through the pipeline.
The photometry pipeline is described in \citet{Tang2013}, whose results can be accessed in a database on their website.
\footnote{Their website with the access to their photometry can be found under \url{http://dasch.rc.fas.harvard.edu/lightcurve.php}.}
The plates from the pipeline have photometric errors that range from 0.04 mag to 0.4 mag with a median error of 0.14 mag, none of which show a sign of an ongoing eclipse.

Eighteen plates from the pipeline were taken with a red or yellow filter, recorded pairwise at identical epochs with plates with a blue filter to form  RYB-triplets at single epochs.
Since a single observation at an arbitrary epoch suffices to exclude a test period (see Section \ref{sect_analysis} for a detailed description), and all plates not accessible via the pipeline were taken with a blue filter, we restrict our photometric analysis of the photographic plates to those taken with a blue filter.

We analysed the other set of plates visually to confirm either the existence or the absence of a transit, and rejected 335 plates which failed to provide reliable photometry of J1407, leaving 52 plates for further analysis.
The visually inspected plates consisted of $\sim 100$px $\times$ 100px sections with J1407 at its centre, which were too small to conduct a colour correction and automated photometry.
A well recorded photographic plate from a sufficient aperture (>4cm) usually has a sufficient sensitivity to  detect a drop in brightness of 0.5 magnitudes.
In several cases, artefacts like blurred and malformed stars or the sky noise prevented an accurate flux measurement of J1407.
The remaining plates were used for the search of possible orbital periods.

\begin{figure}
\centering
\includegraphics[width=1.0\linewidth]{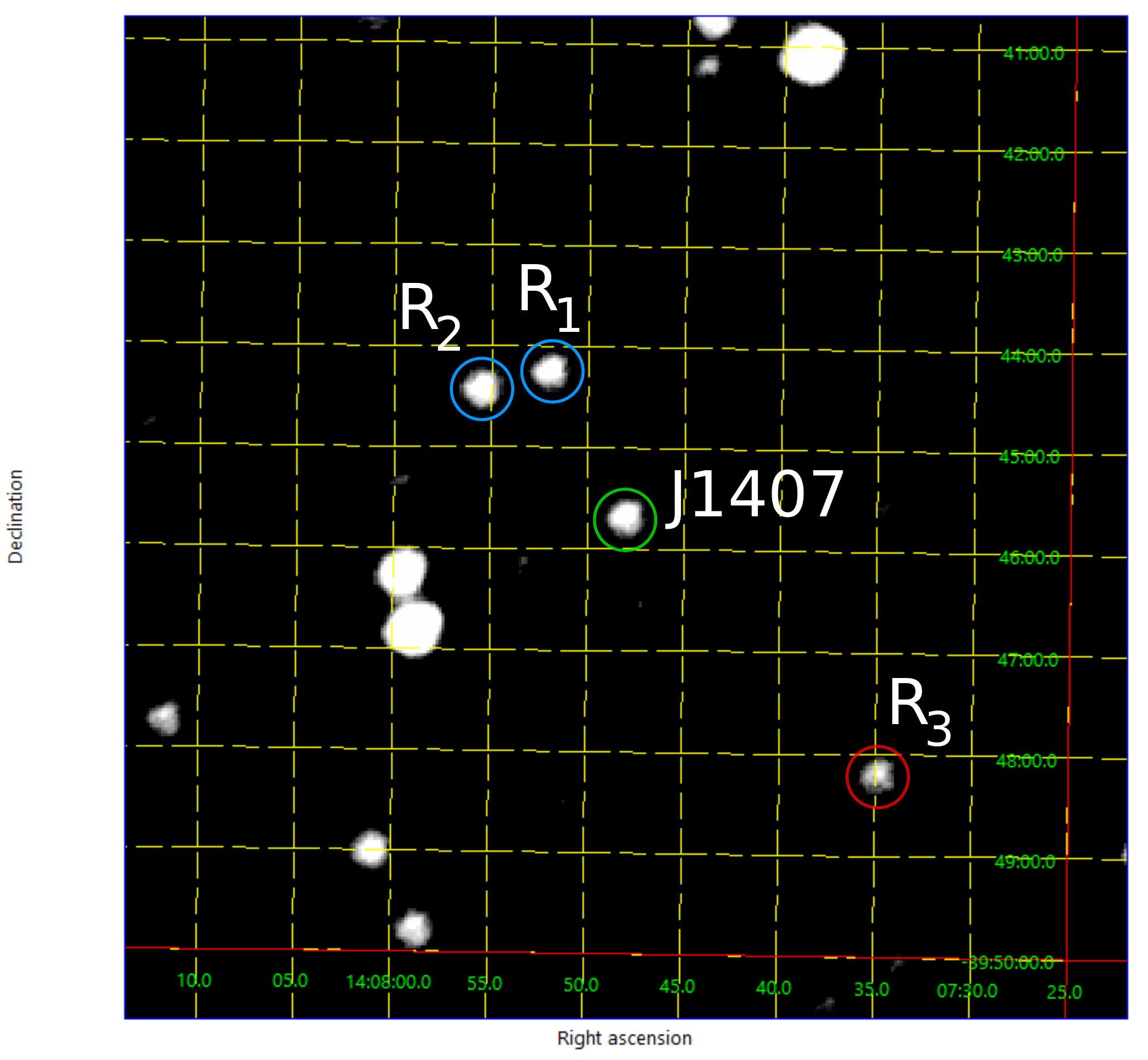}
\caption{Finding Chart for J1407. Green circled: J1407. Blue: Bright Reference stars $R_1$ and $R_2$. Red: Dim reference star $R_3$. The field of view is approximately 10'x10'. The plate depicted (`1948.4914\_mf37404\_00') stems from the `mf'-series from the DASCH archive and was recorded at 1948-06-28.}
\label{fig_find_chart}
\end{figure}

We compared the brightness of J1407 to (a) a region of background sky to estimate the noise, (b), two nearby reference stars $R_1$ and $R_2$ whose apparent magnitudes $m_B$ are both similar to $m_B$ of J1407 out of eclipse, and (c) to a reference star $R_3$ whose $m_B$ is comparable to $m_B$ of J1407 during the 2007 eclipse (see Figure~\ref{fig_find_chart} and Table ~\ref{tab:refstars}).
If J1407 appears to be significantly dimmer than $R_1$ and $R_2$ and not brighter than $R_3$, it can be assumed that the photographic plate was taken while the ring system was transiting.
However, no {\it bona fide} plate that showed such a behaviour was found.
If J1407 appears to be significantly brighter than $R_3$ and at least as bright as $R_1$ and $R_2$, we confidently can rule out a transit of the ring system.
In the other 335 cases, we were not able to confidently either rule out or confirm the photometry of J1407 due to low signal to noise on the plate. This leaves 636 recordings from the DASCH archive for our period searching algorithm.

\subsection{Photographic plates from Bamberg Observatory}

The 171 photographic plates from the Bamberg Observatory were taken between 1964 and 1976.
We obtained photometry for 129 of them using the APPLAUSE pipeline\footnote{\url{https://www.plate-archive.org/applause/}}.
The pipeline did not include photometry for the other 42 plates.
All plates were recorded in Bloemfontein, SA, with one of six 10cm Tessar $f/6$ cameras.
The errors range between 0.17 mag and 0.35 mag with a median of 0.23 mag.
20 of these plates were rejected because of poor photometry, which in total left 109 observations for use in our period folding algorithm.

\subsection{Photographic plates from Sonneberg Observatory}
There are 103 photographic plates covering J1407 during the years of 1935 and 1938, and between 1952 and 1953 in the photographic plate archive at the Sonneberg Observatory (representing the largest European astronomical photographic archive).
Although the Sonneberg Observatory is located in the northern hemisphere, there are a number of southern sky photographic plates and films.
These were taken during the expeditions of the former Director of the Observatory, Professor Cuno Hoffmeister, to South Africa.
These expeditions took place between November 13, 1934 and February 3, 1938, as well as between August 6, 1952 and July 14, 1953.
The Ernostar camera (aperture 135 mm, $f = 24$ cm, $f/1.8$) was used for the observations between 1934 and 1938, while two Tessars ($f=165$ mm, $f/3.5$) were used for the observations in 1952 and 1953.
All instrumentation was positioned at or near Windhuk, Namibia.
The main goal was to monitor selected regions on the southern sky for detection and investigation of variable stars.

The brightness of the star was estimated by an experienced observer (RH) equipped with a magnifying microscope and with the help of nearby constant comparison stars of known magnitudes.
We refer to this as the modified Argelander method.
Our previous investigations confirmed that this method, if carried out by an experienced observer, can provide accuracy comparable with, and under certain circumstances (e.g. if the position is close to the plate edge and/or star images are affected by coma) even better than that of a measuring machine \citep{Hudec76}.
See \citet{Hudec13} for more details on the magnitude estimate procedure.

The star was found to be nearly constant and at its bright stage around $B=13.6$ on all the investigated plates taking into account the measurement errors.
J1407 was close to the magnitude limit on most of the plates so the typical measurement error is $\pm 0.3$ mag. 
We chose a total of 868 archival recordings of J1407 to use in our period folding algorithm.

\subsection{CCD data from KELT}
The Kilodegree Extremely Little Telescope \citep[KELT;][]{Pepper07,Pepper12} survey uses two 42mm telephoto lenses to search for transiting hot Jupiters around bright $(V<11)$ stars.
KELT-South operates at the South African Astronomical Observatory in Sutherland, South Africa.
The telescopes observe the sky with a 10 to 30 minutes cadence, depending on their observing strategy.
Each telescope has a 26 degree square field of view with a 23 arcsecond pixel scale, and can observe stars down to a magnitude $V\sim 14$ at a photometric error of $\sim 4\%$ \citep{Pepper07,Pepper12}.
\footnote{The photometry can be obtained at \url{https://exoplanetarchive.ipac.caltech.edu/cgi-bin/TblSearch/nph-tblSearchInit?app=ExoTbls&config=kelttimeseries}.}
KELT-South has observed J1407 a total of 5534 times between 12.03.2010 and 03.02.2015 (See Figure~$\ref{fig_phot}$).
All these observations show an photometric error around 0.04 mag consistent with the sensitivity of the cameras.
They show no sign of a second transit.

\subsection{CCD data from AAVSO}
In 2012, the American Association of Variable Star Observers (AAVSO)\footnote{\url{http://www.aavso.org/}} started an observing campaign and published an alert to monitor J1407.
The campaign is currently ongoing.
Since then, amateur astronomers have contributed a total of 4342 observations from various sites with various telescopes showing errors between $<0.05$ mag and 0.3 mag which were derived with standard data reduction techniques.
A subset of 99 observations were undertaken either visually or with other filters than Johnson V and will not be used subsequently, leaving 4243 observations for analysis.
They do not show any sign of a transit in the light curve.

\subsection{CCD data from SuperWASP}

The Super Wide Angle Search for Planets \citep{Butters10,Pollacco06} is a robotic observatory undergoing an all-sky survey of the southern (SuperWASP South) and northern hemisphere (SuperWASP North).
Both of these observatories consist of a $2\times4$ array of wide angle
cameras on a common mount with overlapping fields of view.
The southern telescope in Sutherland started surveying J1407 in 2006
and, apart from the ASAS (All Sky Automated Survey), is the only
observer of the transit in 2007.
However, since the ASAS observed the star only a small number of times during the transit, we do not use the ASAS data for our algorithm.
J1407 lies within the overlap region of three single cameras, the 2007 transit was simultaneously observed by these three different cameras.
\citet{Mamajek12,vanWerkhoven14} extensively discussed the photometry from the telescope, including the photometry of the transit.
In total, SuperWASP has observed J1407 close to 29000 times between 05.05.2006 and 09.09.2008.
These observations are available in the SuperWASP public archives\footnote{https://wasp.cerit-sc.cz/form}.

\subsection{CCD data from PROMPT}
The Panchromatic Robotic Optical Monitoring and Polarimetry Telescopes (PROMPT) network of telescopes was constructed by the University of North Carolina at Chapel Hill \citep{Reichart05}.
Part of the network consists of several telescopes located at the Cerro-Tololo Inter-American Observatory (CTIO)\footnote{http://www.ctio.noao.edu/noao/}.
Each telescope is configured differently, allowing for users to pick and choose their observations carefully.
The telescopes use a submission system that automatically queues observations when requested.

PROMPT-4 and PROMPT-5 at CTIO have been observing J1407 in the Johnson V band since mid-2012.
2013 observations have been taken with various configurations of 3 second exposures.
The FITS images are downloaded and processed using AstroImageJ\footnote{http://www.astro.louisville.edu/software/astroimagej/}.
Each image is scanned individually for bad astrometry, hot pixels, and images where clouds moved in or the telescope slewed mid-exposure.
Some of the bad astrometry images were salvageable via plate solving to the WCS\footnote{http://astrometry.net/}.
Hot pixels were corrected using a local median smoothing routine in the software.
Images where clouds moved in and/or the telescope was slewing were removed.
The remaining images are aligned using apertures centered on J1407 and the reference stars.
Multi-aperture differential photometry was performed and fluxes were calibrated to the reference star values from \citet{Kenworthy15}.
The resulting mean uncertainty of the star's brightness is 0.018 mag.

\begin{table*}[t]
\centering
\caption{Brightness and position of the reference stars used in the visual examination of the photographic plates.}
\label{tab:refstars}
\begin{tabular}{ccccc}
\hline
Star Name & Mag in B & Mag in V & Position & Designation \\
\hline
$J1407$ & 13.56 & 12.357 & 14:07:48.36 -39:45:41.9 & 1SWASP J140747.93-394542.6 \\
$R_1$ & 13.601 & 12.413 & 14:07:52.00 -39:44:14.6 & 1SWASP J140752.03-394415.1 \\
$R_2$ & 13.668 & 13.184 & 14:07:55.29 -39:44:26.3 & UCAC2 14835486 \\
$R_3$ & 14.113 & 13.42 & 14:07:34.89 -39:48:12.0 & 2MASX J14073452-3949066 \\
\hline
\end{tabular}
\end{table*}

\begin{figure*}[!ht]
\centering
\includegraphics[width=1.2\textwidth,angle=90]{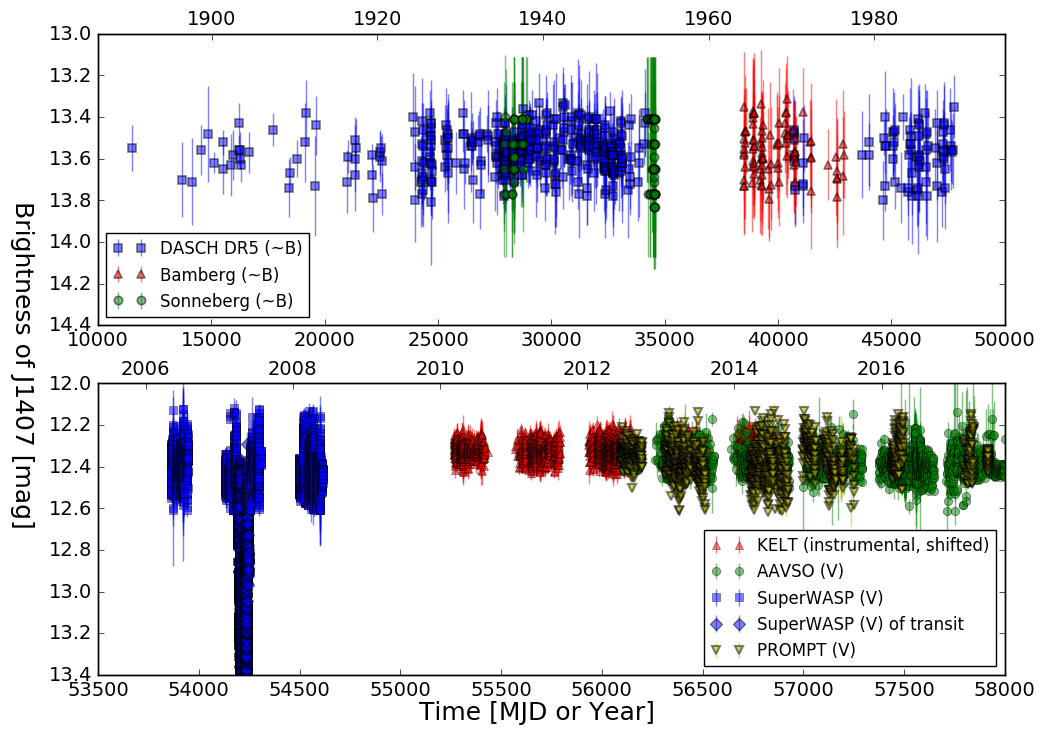}
\caption{Photometry of the data from four CCD sources as well as three photographic plate archives. The offset of the two y-axes between the two separate plots emerges from the different filters with which the observations were taken.}
\label{fig_phot}
\end{figure*} 

\section{Analysis}
\label{sect_analysis}
\subsection{The Period Folding Algorithm (PFA)}
We wrote an algorithm to determine the set of possible orbital periods which are not precluded from a number of observations that bona fide do not show a second transit.
An outline of the procedure is as follows:
The photometric series of J1407 is represented as $m(t)$ where $m$ is the apparent magnitude of J1407 (possibly at different wavelengths respectively in different filters) at epoch $t$.
It is hypothesized that the dimming of the star in the optical due to a transit of the ring system is sufficiently independent of the observed wavelength.
The eclipse in 2007 was detected in Johnson V and the plates used in our analysis were taken in Johnson B.
Small dust particles will preferentially scatter and/or absorb blue (shorter wavelength) light, and so the B plates are more sensitive than the V band observations in this case.
In the `worst' scenario of the dust consisting of larger particles, the absorption will be approximately wavelength independent.
A grey eclipse is therefore a relatively conservative assumption.
For further analysis, we will not take into account the photometry of J1407 since the PFA only requires the value of the epoch $t$ of those observations, which with a reasonable certainty do not show an eclipse.
To exclude a set of orbital periods of J1407b the whole data set is folded into a test period $P$ such that the beginning of the 2007 transit $t_0$ marks the beginning of the phased light curve, so that a plot of $m(t)$ versus $t_{fold}$, where $t_{fold} = (t - t_0) \mod P$, will show any additional eclipses at period $P$ in the range of $t_{fold}$ from 0 to 52 days (which out of 56 days presents the significant part of the total length transit of 2007).

\begin{figure}
\centering
\includegraphics[width=\linewidth]{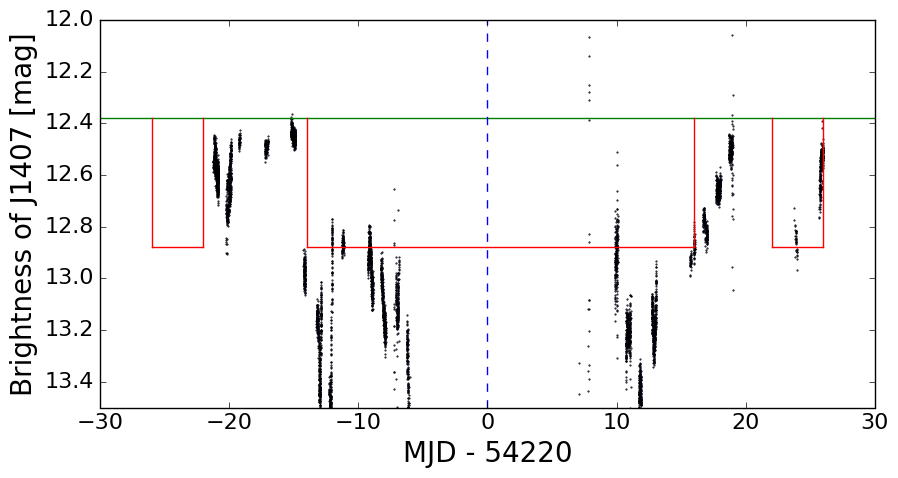}
\caption{Lightcurve of the 2007 transit from the SuperWASP-South observatory (black points). The eclipse is approximated with three windows (red line) corresponding to the three major dimming events. The centre of the transit is marked with a dashed blue line at around MJD 54220.}
\label{fig_model_lightc}
\end{figure}

\begin{figure}
\centering
\includegraphics[width=\linewidth]{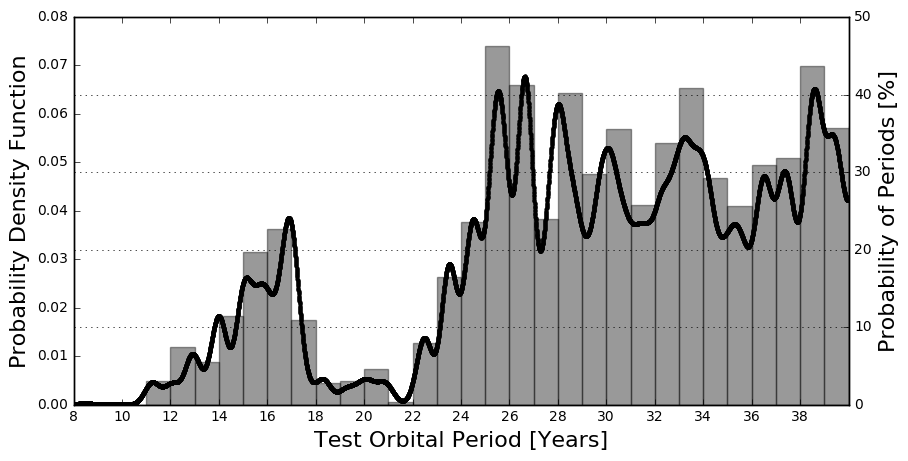}
\caption{Percentage of possible orbital periods of J1407b inferred from our Period Folding Algorithm with 33 steps per day. Kernel Density Estimator (KDE) of the set of possible orbital periods with a Gaussian Kernel and a bandwidth of 0.04.}
\label{fig_prob_plot}
\end{figure}

The 2007 eclipse is approximated as a boxcar function consisting of three transit windows where the photometry was more than 0.5 magnitudes fainter than the out of transit flux levels, as seen in Figure \ref{fig_model_lightc}.
The central eclipse lasts for 30 approximately days from MJD 54204 to 54234, with the second and third windows lasting 4 days from MJD 54194 to MJD 54298 as well as from MJD 54242 to MJD 54246.
For each given trial period $P$, we count the number of observations within the three transit windows $n_{transit}$.
If $n_{transit}=0$ for the test period $P$, it is declared a possible orbital period.
If $n_{transit}>0$, the mean magnitude of these points is calculated.
If the magnitude of the in transit points is not statistically significant, then we can rule out the test orbital period $P$, since the photometry of that epoch should exhibit a sign of a transit.
We then compute the percentage of possible periods in bins of one year each and normalize the distribution per bin to 100\%.
We search periods over a range $0<P<40{\rm yr}$ with 33 steps per day (corresponding to 12175 steps per year) and plot the resulting binned probabilities in Figure~\ref{fig_prob_plot}.
This number of steps was chosen to be sure that the space between the steps is smaller (about 45 minutes) than the uncertainty of the recording time of most photographic plates (about an hour).

\subsection{Results}
We detect no significant transits of J1407 in the data.
The PFA rules out about 90\% of the possible periods from 10 yr to 20 yr.
The probability of the trial periods rise to a peak of 20\% at 16 years, falling to almost zero at 21 years.
As can be seen in Figure \ref{fig_phot}, a higher-than-average number of plates were recorded in the late 1980s (DASCH), as well as early 1970s (Bamberg), and 1950s (Sonneberg).
The folding of these data points thus excludes a great number of possible periods between 18 years and 22 years.
The percentage rises again to 25 years and then stays constant until 40 years of orbital period at values around 25$\%$ to 40$\%$.
Since the average density of data points in the Period Folding inversely scales with the test period, we expect the resulting percentage of possible periods to constantly rise for test periods of 30 years and more.
We cannot confidently rule out any period above 25 years.
Since very few plates have been taken before 1900 - 107 years before the 2007 transit - this method effectively cannot rule out any periods above 100 years.
We expect the next transit no earlier than Spring 2018 (corresponding to an orbital period of 11 years) and most plausibly between Spring 2021 (orbital period of 14 years) and Spring 2024 (orbital period of 17 years).

Analysis using histogram plots is susceptible to the precise placement of the bin edges.
To mitigate this sensitivity, a method called Kernel Density Estimation (KDE) uses a smooth, continuous function (the kernel) to convolve with the data and reduce sensitivity to bin edge placement.
We apply KDE to the data and plot the results in Figure~\ref{fig_prob_plot}. 
We used a Gaussian Kernel and a standard Scott \& Silverman bandwidth selection for the subsets of periods below 20 years and a bandwidth of 0.04 for the whole data set.
The later value was used to have a meaningful depth of details over the entire course of the plot.
The results strongly constrain the orbital period to a value between 11 and 18 years and values above 22 years.
As the probability density rises from 11 years onward, more probable periods range between about 14 years to 17 years with the most probable period around 16.5/17 years.
The peaks of the density function at or close to integer year periods have the following origin.
Since J1407 is not a circumpolar star, we lack photometric data for more than 50\% of the year.
For integer test periods, a number of these gaps in the data thus have about the same phase.
If these gaps now happen to overlap with the test windows of the PFA, it is likely that the PFA yields a higher density of possible periods, hence the peaks seen at integer year spacings.
If our photometry was distributed homogeneously across the course of the year, this feature would be absent.
This phenomena is also present for periods higher than 20 years as can be seen in Figure~\ref{fig_prob_plot}.
The KDE used is the class {\tt scipy.stats.gaussian\_kde} implemented with {\tt scipy}.

\section{Discussion}
\subsection{Circular orbits}

For a given orbital period $P$ and a mass of the secondary companion $m_b$, constraints of the size of its Hill sphere can be derived.
Together with the transverse velocity of the companion and the duration of the 2007 eclipse, we calculate the size of the putative ring system in units of the Hill sphere diameter, $\xi$.
Given Keplerian circular orbits for J1407b, any period greater than 14 to 15 years is improbable considering they cannot explain the high orbital velocity derived from the rate of change of the light curve \citep{Mamajek12,Kenworthy15}.
The coplanarity of a giant ring system with a significant obliquity was examined in \citet{Zanazzi17}, and shown that even for circular orbits, it is impossible to keep ring systems aligned with the Laplacian plane of the planet - at a large enough radius, the obliquity of the rings will decrease to zero.
It therefore seems highly unlikely that J1407b is on a circular and bound orbit.

\subsection{Elliptical orbits}

The rapid variations in the 2007 light curve can be explained by a highly elliptical orbit with the transit occurring close to or during periastron.
Since a longer orbital period infers a higher ellipticity, the minimum orbital period of 11 years only allows for highly elliptical solutions of $0.72<e<0.78$ \citep{Kenworthy15}.
\citet{vanWerkhoven14} have argued that longer periods are unlikely due to the disruption of the ring system during periastron passage, if the constraint of the large transverse velocity derived from the light curve is kept.
Combined with the general constraints on the orbital period from this paper, these dynamical arguments render orbital periods of more than 18 years very unlikely.
If the orbital period is from 11 to 14 years, there is the possibility that the companion can have a mass up to 80 Jupiter masses \citep{Kenworthy15}, although this has a low probability as seen in Figure \ref{fig_prob_plot}.
If the orbital period is from 14 to 18 years, we can then constrain the mass of the ringed companion to be 5 to 20 Jupiter masses with $\xi<1$ for the majority of the orbit.

However at periastron passage, the instantaneous $\xi$ is greater than one causing rings to spill out of the Hill sphere.  \\
\citet{Rieder16} explored the full parameter space of secondary companion mass, ellipticity and periodicity. \\
The conclusion was that prograde orbiting rings were not stable enough through periastron passage to not lose a significant fraction of the outer rings, but that retrograde orbits for the ring particles allow for significantly larger rings out to 60\% of the Hill sphere radius and significant survival of the outer rings through periastron.
Adding retrograde orbits further deepens the challenge of explaining the nature and stability of the ring system seen towards J1407.

\section{Conclusions}\label{sec:conc}

In this paper we have presented photometry from 868 photographic plates from 1890 to 1990, and have combined them with more recent epoch photometry to constrain the orbital period of J1407b.
We detect no statistically significant dimming of the star in all the photographic plates.
We can rule out 90\% of all periods between 10 to 20 years, falling rapidly to a minimum from 18 to 22 years.
Elliptical orbits with periastron passage during May 2007 allow a range of possible masses from 5 to 20 $M_{Jup}$ and infer an ellipticity of about 0.75.
In this case, the rings would fill 70 to 100\% of the Hill sphere of companion, and exceed the shrinking hill sphere during periastron by at least a factor of two.
Although the range of orbital periods has been reduced, the lack of the detection of another eclipse means that we still cannot confirm that the eclipse event in 2007 was due to an extended object on a bound orbit around J1407.

Two possible future observations that can resolve the nature of the J1407 eclipse include continued photometric monitoring for the next eclipse, and direct detection of radiation from the rings themselves, either at sub-mm wavelengths with ALMA, or in optical wavelengths through the polarization of reflected light from the star.

\begin{acknowledgements}
We express our deepest gratitude to Alison Doane, former Curator of the Harvard Astronomical Plate collection, who sadly passed away during the course of this research.
Her expansive knowledge of the history of the Harvard Plates and the Harvard women computers who studied them, and her gracious patience in instructing us on how best to use the plates was instrumental to this research.
Her quick action during the flood of Harvard College Observatory in January 2016 was vital in saving the photographic plates and preserving this unique historical resource for many future generations.
Part of this work was supported by the DAAD programme allowing RTM to carry out the first part of this research in Leiden Observatory during the Summer of 2016.
      MAK thanks DASCH for allowing access to their archives.
      The DASCH project at Harvard is grateful for partial support from NSF grants AST-0407380, AST-0909073, and AST-1313370.
      Funding for APPLAUSE has been provided by DFG (German Research Foundation, Grant), the Leibniz Institute for Astrophysical Research (AIP), the Dr. Remeis Sternwarte Bamberg (University Nuernberg/Erlangen), the Hamburger Sternwarte (University of Hamburg) and Tartu Observatory.
      RH thanks Sonneberg Observatory for access to its plate archive and acknowledges GACR grant 13-33324S.
      We acknowledge with thanks the variable star observations from the AAVSO International Database contributed by observers worldwide and used in this research.
This research made use of {\tt Astropy}, a community-developed core Python
package for Astronomy \citep{2013A&A...558A..33A}.
This work has made use of data from the European Space Agency (ESA) mission
{\it Gaia} (\url{https://www.cosmos.esa.int/gaia}), processed by the {\it Gaia}
Data Processing and Analysis Consortium (DPAC,
\url{https://www.cosmos.esa.int/web/gaia/dpac/consortium}). Funding for the DPAC
has been provided by national institutions, in particular the institutions
participating in the {\it Gaia} Multilateral Agreement.
SNM is a U.S. Department of Defense SMART scholar sponsored by the U.S. Navy through SSC-LANT.
This work was performed in part under contract with the Jet Propulsion Laboratory (JPL) funded by NASA through the Sagan Fellowship Program executed by the NASA Exoplanet Science Institute.
E.E.M. and S.N.M. acknowledge support from the NASA NExSS programme. 
Part of this research was conducted at Jet Propulsion Laboratory, California Institute of Technology, under a contract with the National Aeronautics and Space Administration.
    We thank the referee for their time and constructive comments that have improved this paper.

\end{acknowledgements}

\bibliographystyle{aa}
\bibliography{mybib}

\end{document}